\documentclass{PoS}

\title{Impact of oscillations of photons into axion-like particles on the very-high energy {\gray} spectrum of the blazar PKS\,1424+240}

\ShortTitle{Impact of ALPs on PKS\,1424+240}

\usepackage{thesis}

\author{\speaker{Manuel Meyer}\\
	Oskar Klein Centre, Department of Physics, Stockholm
	University, Albanova University Center, SE-10691
	Stockholm, Sweden\\
	\emph{and}\\
	Universit\"at Hamburg, Institut f\"ur Experimentalphysik,
	Luruper Chaussee 149, D 22761 Hamburg, Germany
        \\
        E-mail: \email{manuel.meyer@fysik.su.se}}

\author{Dieter Horns\\
	Universit\"at Hamburg, Institut f\"ur Experimentalphysik,
	Luruper Chaussee 149, D 22761 Hamburg, Germany
        \\
        E-mail: \email{dieter.horns@desy.de}}

\abstract{
Very high energy (VHE) gamma-rays undergo pair production
with low energy photons of background radiation fields. 
This leads to an attenuation of the primary gamma-ray flux of extragalactic sources
in the interaction with the extragalactic background light (EBL)
which stretches from ultraviolet to far-infrared wavelengths.

In the presence of magnetic fields, gamma-rays could oscillate into hypothetical
axion-like particles (ALPs). 
This might lead to a reduced opacity of the Universe for VHE gamma-rays,
as ALPs circumvent pair production.
Here, the impact of photon-ALP conversions on the spectrum 
of PKS\,1424+240 is demonstrated. A lower limit on the 
redshift of this blazar was recently determined to be $z \geqslant 0.6035$,
making it the farthest source ever observed at VHE energies.
Under the assumption of a specific magnetic field scenario and 
EBL model, photon-ALP couplings are derived that lead to an overall
concave intrinsic blazar spectrum.
}

\FullConference{The European Physical Society Conference on High Energy Physics -EPS-HEP2013\\
		18-24 July 2013\\
		Stockholm, Sweden}

\begin{document}

\section{Introduction}
\begin{figure}[thb]
 \centering
 \includegraphics[width = 0.65\linewidth]{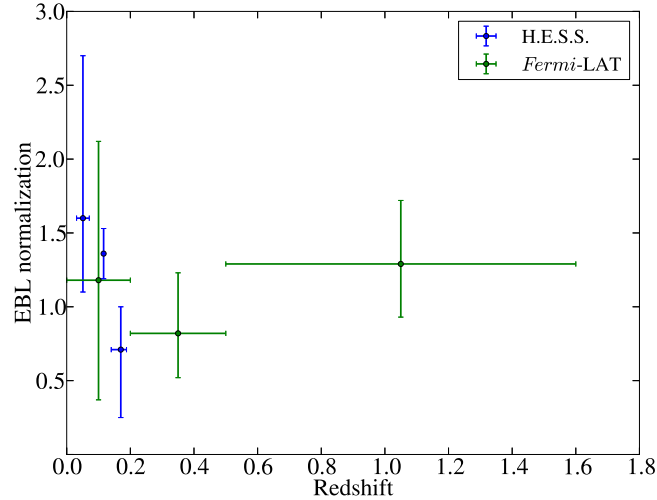}
 \caption{Best-fit EBL normalization determined by \cite{ackermann2012,hess2013ebl}. Both studies used the EBL model of Ref. \cite{franceschini2008}.}
 \label{fig:ebl-det}
\end{figure}

The interaction of {\grays} with low energy photons originating from the extragalactic background light (EBL), $\gamma\gamma_\mathrm{EBL} \to e^+e^-$,
leads to an exponential suppression, $\exp(-\tau)$, of the {\gray} flux emitted by an extragalactic source at high and very high energies 
(VHE, $E \gtrsim 100$\,GeV; see Ref. \cite{dwek2013} for a review).
Recently, the imprint of the absorption was detected in 150 blazar spectra
with the \emph{Fermi}-LAT within a redshift range $0 < z < 1.6$ \cite{ackermann2012}. 
The effect was independently observed in data from mainly two blazars observed with H.E.S.S. \cite{hess2013ebl}. 
Both studies used model templates for the EBL and best-fit values for the normalization of the 
optical depth were derived. The results are summarized in Fig. \ref{fig:ebl-det}, showing the normalization for different redshift bins.
The measurements are consistent with each other within their uncertainties and indicate a decrease of the opacity towards higher redshifts. 
Additionally, a multiwavelength modeling of spectral energy distributions (SEDs) of blazars 
shows clear evidence for the expected dimming of the VHE flux \cite{dominguez2013}.

Any modifications to this picture by a non-standard propagation of photons are likely to become competitive 
at high optical depths, $\tau > 2$. 
Indeed, several studies found indications for a low opacity of the Universe for VHE {\grays} 
\cite{1es0229hess2007,3c279magic2008,essey2012,deangelis2011,horns2012,meyer2012ppa},
and also in observations with the \emph{Fermi}-LAT \cite{horns2013,meyer2013thesis}.
These findings have been interpreted in several scenarios, 
e.g., the efficient acceleration of protons by blazars 
that subsequently interact with background radiation fields and initiate electromagnetic cascades \cite{essey2010}.
The produced secondary photon flux enhances the flux observed on Earth.
Alternatively, photons can evade pair production if they oscillate into hypothetical axion-like particles (ALPs) in ambient magnetic fields. 
The conversion of VHE {\grays} into ALPs has been studied in different magnetic field settings, 
including the intergalactic magnetic field (IGMF, see Refs. \cite{deangelis2007,mirizzi2007}), intracluster magnetic fields (ICMF, see Ref. \cite{horns2012ICM}),
the Galactic magnetic field (GMF, see Ref. \cite{simet2008}), and the magnetic field in the source itself \cite{hochmuth2007}.
In Ref. \cite{meyer2013} different magnetic field scenarios were investigated in order to derive 
lower limits on $\gag$ under the assumption that ALPs are responsible for the indication of a reduced opacity at VHE.
Furthermore, photon-ALP oscillation could lead to irregularities in spectra \cite{ostman2005,wouters2012} as well as distinct spectral features around the energy where 
the mixing becomes maximal \cite{hochmuth2007}.

Here, the effect of ALPs on the VHE spectrum of the blazar PKS\,1424+240 is studied.
The BL Lac object was observed with VERITAS in 2009 
and its spectrum is best described with a simple power law, $\Difft{N}{E} \propto E^{-\Gamma_\mathrm{VHE}}$,
with $\Gamma_\mathrm{VHE} = 3.8\pm0.5_\mathrm{stat} \pm0.3_\mathrm{sys}$ \cite{pks1424veritas2010}.
No indication for variability was found during the observations with neither VERITAS nor the \emph{Fermi}-LAT. 
Between 100\,MeV and 300\,GeV, the analysis of the \emph{Fermi}-LAT data yielded a spectral index of $\Gamma_\mathrm{Fermi} = 1.73\pm0.07_\mathrm{stat} \pm0.05_\mathrm{sys}$.
Recently, the a firm lower limit on the source redshift could be determined from optical observations
through the detection of intervening Ly-$\beta$ and -$\gamma$ absorption lines, $z \geqslant 0.6035$ \cite{furniss2013}.
This makes the source the most distant blazar observed at VHE with an optical depth 
$\tau \gtrsim 5.5$ for the highest energy data point at $\sim$\,470\,GeV (utilizing the best-fit optical depth normalization and EBL model determined in Ref. \cite{hess2013ebl}).
Thus, the expected absorption is severe and photon-ALP oscillations can, in principle, lead to a significant boost of the apparent photon flux.
In the following, the VERITAS spectrum will be corrected for EBL absorption with and without the effect of ALPs. 
The resulting spectrum will be compared to the \emph{Fermi}-LAT spectrum in the energy range where the attenuation is negligible.
Under the assumption of the best-fit EBL model of \cite{hess2013ebl} and that the intrinsic blazar spectrum does not harden towards higher energies, 
it is possible to determine the minimal photon-ALP coupling for which $\Gamma_\mathrm{VHE} \geqslant \Gamma_\mathrm{Fermi}$,
i.e., the overall absorption-corrected {\gray} spectrum is concave, as expected in simple blazar emission models\footnote{
Although not expected in the simplest blazar models, spectral hardening can occur 
as demonstrated in specific scenarios (see, e.g., Ref. \cite{aharonian2008}). 
Furthermore, the usage of the \emph{Fermi}-LAT spectrum as an estimator for the intrinsic spectrum is debated \cite{costamante2012}.
}.

\section{Absorption-corrected {\gray} spectrum of PKS\,1424+240}
\begin{figure}[thb]
 \centering
 \includegraphics[width = 0.65\linewidth]{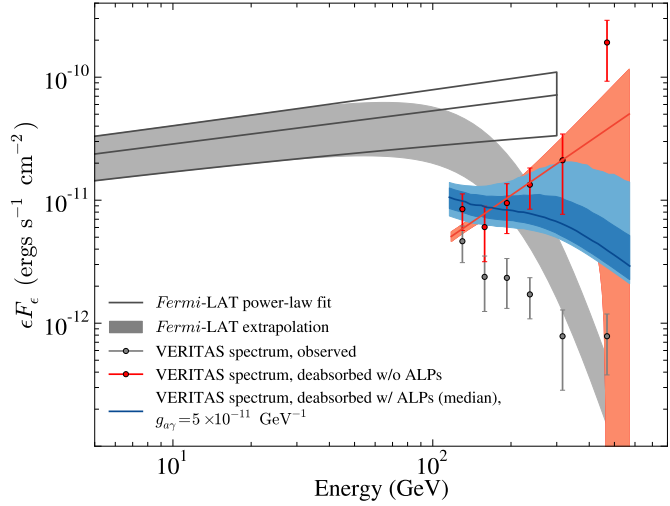}
 \caption{Spectral energy distribution for PKS\,1424+240. The gray-shaded area shows the extrapolation of the \emph{Fermi}-LAT
 spectrum under the assumption of the best-fit EBL model determined in Ref. \cite{hess2013ebl}.}
 \label{fig:sed}
\end{figure}

The intrinsic spectrum of PKS\,1424+240 can be estimated from \emph{Fermi}-LAT measurements contemporaneous with VERITAS observations
that yielded a power law with  spectral index of $\Gamma_\mathrm{Fermi} = 1.73 \pm 0.07_\mathrm{stat} \pm 0.05_\mathrm{sys}$ between 100\,MeV and 300\,GeV \cite{pks1424veritas2010},
and is shown in Fig. \ref{fig:sed}.
Above $\sim27\,$GeV, the attenuation becomes larger than 1\,\%, thus, the \emph{Fermi}-LAT spectrum already extends into the energy regime where  absorption becomes significant. 
However, this will mainly affect the high energy end of the spectrum where count statistics are low and the impact on the overall spectrum should be small. 

The figure also shows the observed and absorption-corrected data points of the VERITAS observation (gray and red bullets, respectively).
A fit to the absorption-corrected data points results in a spectral index of $\Gamma_\mathrm{VHE} = 0.55\pm0.56$, in conflict with
the assumption that $\Gamma_\mathrm{VHE} \geqslant \Gamma_\mathrm{Fermi}$, even if the statistical uncertainties are taken into account. 
This remains true even if the EBL normalization $\alpha$ is scaled within the statistical uncertainties derived in \cite{hess2013ebl} 
from $\alpha = 1.27$ to $\alpha - \sigma_\alpha = 1.12$. This normalization gives a softer spectral index of $\Gamma_\mathrm{VHE} = 0.85\pm0.56$.

The effect of ALPs is demonstrated within a specific scenario (the same scenario labeled \textit{fiducial} in Ref. \cite{meyer2013},
where more details are given):
it is assumed that PKS\,1424+240 is located in the center of a galaxy cluster with radius of 2/3\,Mpc that 
is filled with a cell-like structured magnetic field with a field strength of 1\,$\mu$G. 
The IGMF is modeled with the same morphology and with a field strength of 0.01\,nG.
The large-scale regular component of the GMF is described with the model outlined in \cite{jansson2012}.
In total, 5000 realizations are simulated each for the randomly orientated ICM and IGMF.

\begin{figure}[thb]
 \centering
 \includegraphics[width = 0.65\linewidth]{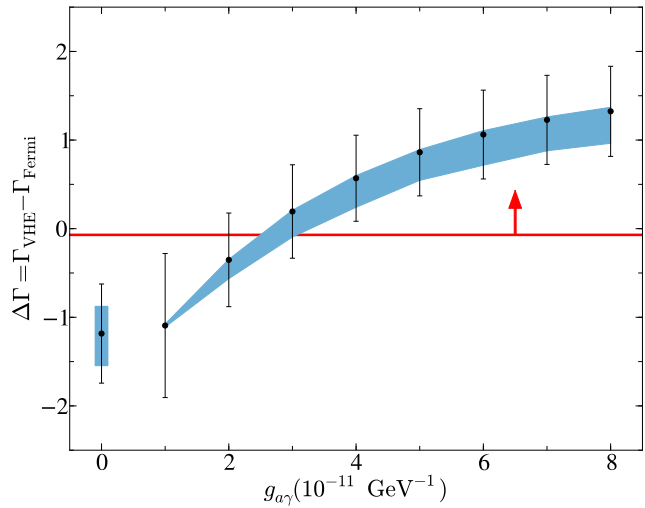}
 \caption{Difference of spectral indices at high and very high energies. Points below the uncertainty of the index determined with \emph{Fermi}-LAT data
 (red solid line) indicate convex spectra not expected in simple blazar models.}
 \label{fig:dgamma}
\end{figure}

The observed VERITAS spectrum corrected with the median of all realizations along with the 68\,\% and 95\,\% confidence contours 
is shown in Fig. \ref{fig:sed} for an photon-ALP coupling of $5\times10^{-11}\,\mathrm{GeV}^{-1}$ and $\ma = 1\,\mathrm{neV}$.
The choice of this particular mass ensures a maximal mixing between photons and ALPs and is assumed henceforth.
Clearly, the resulting intrinsic spectrum is softer than in the no-ALPs case.
The difference between spectral indices, $\Delta\Gamma = \Gamma_\mathrm{VHE} - \Gamma_\mathrm{Fermi}$,
for different values of $\gag$ is shown in Fig. \ref{fig:dgamma}.
For each value, the observed spectrum is corrected for absorption with the median of all realizations. 
If, instead of the median, the 68\,\% confidence contours of all $B$-field realizations (cf. the dark-blue shaded region in Fig. \ref{fig:sed})
are used to calculate the intrinsic spectrum, the best-fit spectral index changes according to the blue-shaded region in Fig. \ref{fig:dgamma}.
For $\gag \gtrsim4\times10^{-11}\,\mathrm{GeV}^{-1}$ the resulting intrinsic index 
is softer than the Fermi index, taking the statistical uncertainties into account (the uncertainty 
of the \emph{Fermi} index is shown as a red solid line in Fig. \ref{fig:dgamma}).
The blue shaded region in case of no ALP mixing ($\gag = 0\,\mathrm{GeV}^{-1}$)
indicates the fit results if $\alpha$ is varied within its statistical uncertainties.

\section{Summary and discussion}

Under the assumption of a specific magnetic field scenario, it has been shown that the conversion 
of photons into hypothetical ALPs can lead to absorption-corrected spectra that do not 
show a spectral hardening as predicted from standard blazar emission scenarios.
This cannot be achieved if the current best fit of the EBL determined from VHE blazar observations is used.
Interestingly, the obtained minimum value of $\gag$ that leads to an overall concave intrinsic spectrum 
agrees well with the lower limits on $\gag$ determined in Ref. \cite{meyer2013}.

Although these findings cannot be directly interpreted as evidence for ALPs, 
they highlight the potential importance of these particles for the propagation of photons. 
Future observations with the \emph{Fermi}-LAT (especially with the upcoming \textit{Pass 8} event class release),
currently operating imaging air Cherenkov telescopes (IACTs) and the 
the Cherenkov Telescope Array (CTA) will shed further light on the opacity of the Universe and 
the existence of ALPs.
The latter will also be tested in future laboratory experiments (e.g., Ref. \cite{alpsII}).


\bibliographystyle{plain}
\bibliography{vhe_spectra,meyer_ALP_transparency}    

\end{document}